\documentclass[10pt, letterpaper]{article}

\let\n\noindent

\newcommand{\beq}{\begin{equation}} \newcommand{\eeq}{\end{equation}}
\newcommand{\beqa}{\begin{eqnarray}}    \newcommand{\eeqa}{\end{eqnarray}}
\newcommand{\btab}{\begin{tabular}}     \newcommand{\etab}{\end{tabular}}

\newcommand{\bt}{\begin{table}}     \newcommand{\et}{\end{table}}

\newcommand{\ba}{\begin{array}}     \newcommand{\ea}{\end{array}}
\newcommand{\bc}{\begin{center}}        \newcommand{\ec}{\end{center}}
\newcommand{\bfig}{\begin{figure}}      \newcommand{\efig}{\end{figure}}
\newcommand{\bp}{\begin{picture}}       \newcommand{\ep}{\end{picture}}
\newcommand{\bq}{\begin{quote}}     \newcommand{\eq}{\end{quote}}
\newcommand{\ben}{\begin{enumerate}}    \newcommand{\een}{\end{enumerate}}

\font\tenmsy=msbm10
\font\sevenmsy=msbm10 at 7pt
\font\fivemsy=msbm10 at 5pt
\newfam\msyfam 
\textfont\msyfam=\tenmsy
\scriptfont\msyfam=\sevenmsy
\scriptscriptfont\msyfam=\fivemsy

\def\frac#1#2{{\textstyle{#1\over #2}}}

\def\text#1{\quad\hbox{#1}\quad}

\def\la{\lambda}
\def\e{\epsilon}
\def\ka{\kappa}

\def\A{{\cal{A}}}
\def\At{{\tilde{A}}}

\def\Bt{{\tilde{B}}}

\def\y{{\infty}}

\def\rw{\rightarrow}

\def\frac#1#2{{#1 \over #2}}

\def\rw{{\rightarrow}}

\begin{document}

\title{Generating function for $K$-restricted jagged partitions
    }

\author{J.-F. Fortin, P. Jacob and P.
Mathieu\thanks{jffortin@phy.ulaval.ca, pjacob@phy.ulaval.ca,
pmathieu@phy.ulaval.ca.  This work is supported by NSERC.} \\ 
\\
D\'epartement de physique, de g\'enie physique et d'optique,\\
Universit\'e Laval, \\
Qu\'ebec, Canada, G1K 7P4.
}

\date{February 2004}

\maketitle


\vskip0.3cm
\centerline{{\bf ABSTRACT}}
\vskip18pt

\n We present a natural extension of Andrews' multiple sums counting partitions with  
difference 2 at distance $k-1$, by deriving the
generating function for  $K$-restricted jagged partitions. 
 A jagged partition is a collection of non-negative integers
$(n_1,n_2,\cdots , n_m)$ with
$n_m\geq 1$ subject to the weakly decreasing conditions $n_i\geq n_{i+1}-1$ and
 $n_i\geq n_{i+2}$. The $K$-restriction refers to the following additional conditions:
$n_i \geq  n_{i+K-1} +1$ or $
n_i = n_{i+1}-1 =  n_{i+K-2}+1= n_{i+K-1}$. The corresponding generalization of the Rogers-Ramunjan identities is
displayed, together with a novel combinatorial interpretation.


\newpage


\section{Introduction}

In 1981 Andrews \cite{Andrr} showed that the generating function for
partitions with prescribed number of parts subject to the following  difference 2 condition
\beq\label{usual}
  \la_j\geq \la_{j+k-1}+2
\eeq
and containing at most $i-1$ parts equal to 1 is 
\beq\label{ande}
F_{k,i}(z;q)= \sum_{m_1,\cdots,m_{k-1}=0}^\y {
q^{N_1^2+\cdots+ N_{k-1}^2+L_{i}}\; z^{N} \over (q)_{m_1}\cdots (q)_{m_{k-1}}
}\;,\eeq
with 
\beq\label{defNL}
 N_j= m_j+\cdots
+m_{k-1}\,,\qquad L_j=N_j+\cdots N_{k-1}\,,\qquad N=L_1\; ,
\eeq
($ L_k=L_{k+1}=0$) and
\beq
(a)_n=(a;q)_n= \prod_{i=0}^{n-1} (1-aq^i)\; .
\eeq
This is a one-parameter deformation of the multiple $q$-series related to the Gordon product \cite{An, Andr}.

In this work, we present the derivation of the generating function  for
jagged partitions of length $m$, which are $m$-component vectors 
$(n_1,n_2,\cdots , n_m)$ with non-negative entries  
satisfying
\beq\label{oror}
 n_j\geq n_{j+1}-1\;,\qquad  \qquad  n_j\geq n_{j+2}\;, \qquad\qquad   n_m\geq 1\;,
\eeq
 further subject to the following $K$-restrictions: 
\beq \label{allo}
n_j \geq  n_{j+K-1} +1 \qquad{\rm or} \qquad
 n_j = n_{j+1}-1 =  n_{j+K-2}+1= n_{j+K-1} \;,
\eeq 
for all values of $j\leq m-K+1$, with $K>2$.  
Following \cite{Andrr}, the derivation of the generating function uses
a  recurrence process  controlled by a  boundary condition, which here is taken to be a constraint on the number of pairs
$01$ that can appear in the $K$-restricted jagged partitions.   Our main result is the following (which is a reformulation of Theorem
7, section 3):

\n {\bf Theorem 1.} If 
 $A_{K,2i}(m,n)$ stands for the set of $m$ non-negative integers $(n_1,\cdots,n_m)$  of
weight $n=\sum_{j=1}^m n_j$ satisfying the weak ordering conditions 
(\ref{oror})
together with the restrictions 
(\ref{allo}) and containing at most $i-1$ pairs 01, then its generating function is
\beq
\sum_{n,m\geq 0} A_{K,2i}(m,n) z^m q^n =  \sum_{m_0,\cdots,m_{\ka-1}=0}^\y { q^{m_0(m_0+1)/2+  \e m_0m_{\ka-1}+
N_1^2+\cdots+ N_{\ka-1}^2+L_{i}}\; z^{m_0+2N} \over (q)_{m_0}\cdots (q)_{m_{\ka-1}} }\;,
\eeq
where $\kappa$ and $\e$ ($=0$ or $1$) are related to $K$ by $K=2\ka-\e$ and where $N_j$ and $L_j$ are given in (\ref{defNL})
with $k$ replaced by $\kappa$. 

Jagged partitions have first been introduced in the context of
a conformal-field theoretical problem
\cite{JM}. In that framework, $K =2\kappa$, i.e., it is an even integer. The generating functions for the $2\kappa$-restricted jagged
partitions with boundary condition specified by $i$  has been found in \cite{BFJM}. It is related to the character of the irreducible
module of the parafermionic highest-weight state specified by a singular-vector condition 
labeled by the integer $1\leq i\leq \kappa$. 

Our essential contribution in this paper is to present the generating function for $K$
odd, for which we do not know any  physical realization.  However, this is a natural mathematical
extension and it turns out  that it is not so straightforward. Moreover, the resulting
generating function has a nontrivial product form, which is given in Theorm 11 (in the even case, the product form
reduces to the usual one in the Andrews-Gordon identity \cite{An}). In all but one case, the resulting
generalizations of Rogers-Ramanujan identities reduce to identities already found by Bressoud
\cite{Bres}.  However, the identity corresponding to $i=\kappa$ (with $K=2\kappa-1$) appears to be
new. But quite interestingly, in all cases (i.e., for all allowed values of $i$ and
$K$, including
$K$ even), we present (in Corollary 12) a new combinatorial interpretation of
 these generalized  Rogers-Ramanujan identities in terms of
jagged partitions.


\section{Jagged partitions}

Let us start by formalizing  and exemplifying the notions of jagged partitions and their restrictions.

\n {\bf Definition 2.} A {\it jagged partition} of length $m$ is a  $m$-component vector
$(n_1,n_2,\cdots , n_m)$ with non-negative entries  
satisfying $n_j\geq n_{j+1}-1, \;   n_j\geq n_{j+2}\; $ and $  n_m\geq 1$.

Notice that even if the last entry is strictly positive, some zero entries are allowed. For instance, the lowest-weight jagged
partition is of the form $(\cdots 01010101)$.  The origin of the qualitative `jagged' is rooted in the
 jagged nature of this lowest-weight vector.  The list of all jagged partitions of length $6$ and weight $7$ is:
\beq \label{case}
\{(410101), \,(320101),\, (230101),\, (311101),\, (221101),\, (212101),\, (211111),\, (121111),\, (121201)\}\;.
\eeq
Observe that to the set of integers $\{0,1,1,1,2,2\}$ there correspond  three jagged partitions of length $6$ and weight $7$ but,
of course, only one standard partition.

\n {\bf Definition 3.} A {\it $K$-restricted jagged partition} of length $m$ is a jagged partition 
 further subject to the conditions: 
$ n_j \geq  n_{j+K-1} +1 $ or $
 n_j = n_{j+1}-1 =  n_{j+K-2}+1= n_{j+K-1} $
(called $K$-restrictions) for all values of $j\leq m-K+1$, with $K>2$.

 The first condition enforces difference 1
 at distance $K-1$. However, the
 second condition allows for some partitions with
  difference 0 at distance $K-1$ if in addition they satisfy
 an in-between 
 difference 2 at distance $K-3$.  In other words, it is equivalent to $ n_j =  n_{j+K-1} $ 
and $n_{j+1} =  n_{j+K-2}+2$.
The general pattern of such $K$ consecutive
  numbers is $(n,n+1,\cdots , n-1,n)$,
 where the dots stand for a sequence of $K-4$ integers 
  compatible with the weak ordering conditions (\ref{oror}).  

The list of all 5-restricted jagged partitions of length $6$ and weight $7$ is
\beq \label{caser}
\{(320101),\, (230101),\,  (221101),\, (212101),\,  (121201)\}\;.
\eeq 
Comparing this list with that in (\ref{case}), we see that $(410101)$ is not allowed since $n_2=n_6$ but $n_3\not=n_5+2$. $(311101)$
and
$(211111)$ are excluded for the same reason. Moreover, $(1211101)$ is excluded since $n_1=n_5$ but $n_2\not=n_4+2$.  $(212101)$
is an example of an allowed jagged partition with an in-between difference 2 condition at distance $K-3=2$.

\section{Recurrence relations for generating functions}

We first introduce two sets of $K$-restricted  jagged partitions with prescribed boundary conditions:

\n $A_{K,2i}(m,n)$:  the number  of $K$-restricted jagged partitions of $n$ into $m$  parts with {\it at most} $(i-1)$
pairs of 01, with $1\leq i\leq [(K+1)/2]$.

\n $ B_{K,j} (m,n)$: the number  of $K$-restricted jagged  partitions of $n$  into $m$  parts with {\it at
most}Ê $(j-1)$ consecutiveÊ$\, 1$'s at the right end, with $1\leq j\leq K$.

\n These definitions are augmented by the specification of the following boundary conditions:
\beq \label{bdryA}
 A_{K,2i}(0,0)= B_{K,j}(0,0) = 1\;,\qquad
 A_{K,0}(m,n)=B_{K,0}(m,n)=0\;.
\eeq
Moreover, it will be understood that both $A_{K,2i}(m,n)$ and $B_{K,j}(m,n)$ are zero when either $m$ or $n$ is negative and if
either of $m$ or $n$ is zero (but not both).

We are interested in finding the generating function for the set $A_{K,2i}(m,n)$. $ B_{K,j} (m,n)$ is thus an auxiliary object whose
introduction simplifies considerably the analysis.

\n {\bf Lemma 4.} The  sets $A_{K,2i}$ and $B_{K,j}$ satisfy the following recurrence relations: 
\begin{eqnarray}\label{recuA} 
&(i)\qquad  & A_{K,2i} (m,n)-A_{K,2i-2 }(m,n)= B_{K,K-2i+2} (m-2i+2,n-i+1)\;,\cr
&(ii)\qquad &B_{K, 2i+1 } (m,n) - B_{K, 2i } (m,n)=Ê A_{K, K-2i+\e} (m-2i,n-m) \;,\cr
&(iii) \qquad & B_{K, 2i } (m,n) - B_{K, 2i-1 } (m,n)=Ê A_{K, K-2i+2-\e} (m-2i+1,n-m) \;,\,
\end{eqnarray} 
where $\e$ is related to the parity of $K$ via its decomposition as
\beq
K=2\ka -\e\qquad (\e=0,1) \,\;. 
\eeq

\n {\it Proof}: The difference on the left hand side of the recurrence relations selects sets of jagged  partitions with a specific
boundary term. In particular, $A_{K,2i} (m,n)-A_{K,2i-2 }(m,n) $ gives the number of
$K$-restricted jagged  partitions of $n$ into $m$  parts containing {\it  exactly} $i-1$
pairs of 01 at the right.  Taking out the tail 
 $01\cdots 01$, reducing then the length of the partition from $m$ to $m-2(i-1)$
and its weight $n$ by $i-1$, we end up with  $K$-restricted jagged partitions which can terminate
with a certain number of 1's. These are elements of the set $B_{K,j}(m-2i+2,n-i+1)$. It remains to fix $j$. The number of
1's in the stripped jagged partitions is constrained by the restriction. Before taking out the tail, the number of
successive 1's  is at most
$K-2(i-1)-1$; this fixes $j$ to be $K-2(i-1)$. We thus get the right hand side of
$(i)$.  By reversing these operations, we can transform elements  of $B_{K,K-2i+2} (m-2i+2,n-i+1)$ into those of
$A_{K,2i} (m,n)-A_{K,2i-2 }(m,n)$, which shows that the correspondence is one-to-one. This proves $(i)$.

Consider now the relation 
$(ii)$.  The left hand side is the number of $K$-restricted jagged  partitions of $n$ into $m$  parts containing {\it  exactly} $2i$
parts equal to  1 at the right end.  Subtracting from these jagged partitions the ordinary partition  $(1^m)= (1,1,1,\cdots,1)$ yields
new jagged  partitions of length $m-2i$ and weight $n-m$.  Since these can have a certain number of pairs of 01 at the end (which is
possible if originally we had a sequence of 12 just  before the consecutive 1's), we recover elements of $A_{K,2i'}(m-2i,n-m)$. It
remains to fix
$i'$. Again, the $K$-restriction puts constraints of the number of allowed pairs 12 in the unstripped jagged partition; it is $\leq
(K-2i+\e-2)/2$. [Take for instance $K=7$ and $2i=4$; the lowest-weight jagged partition of length 7 and four 1's at the
end is
$(2121111)$, which is compatible with the
$6$-step difference-one condition; by stripping off $(1^7)$, it is reduced to $(101)$ so that here there is at most one pair of 01
allowed.  Take instead 
$K=8$ and again $2i=4$; the lowest-weight jagged partition of length 8 is now $(22121111)$, the leftmost 2 being forced by the
$7$-step difference-one condition; it is reduced to $(1101)$ so that here there is again at most one pair of 01 allowed. Note that
for both these examples, the alternative in-between difference-two condition is not applicable.]  Hence $i'=(K-2i+\e)/2=\ka-i$.
Again the correspondence between sets defined by the two sides of $(ii)$ is one-to-one and this completes the  proof of
$(ii)$. The proof of
$(iii)$ is similar.

Let us now define the
generating functions:
\begin{eqnarray}
& \At_{K,2i}(z;q)= \sum_{m,n\geq 0} z^mq^n A_{K,2i}(m,n)\;,\cr
& \Bt_{K,j}(z;q)= \sum_{m,n\geq 0} z^mq^n B_{K,j}(m,n)\;.\,
\end{eqnarray}
In the following, we will generally suppress the explicit $q$ dependence (which will never be modified in our analysis) and write
thus $\At_{K,2i}(z)$ for
$\At_{K,2i}(z;q)$. The recurrence relations $(i)-(iii)$ are now transformed into $q$-difference equations given in the next lemma,
whose proof is direct.

\n {\bf Lemma 5.} The functions $\At_{K,2i}(z;q)$ and $\Bt_{K,j}(z;q)$ satisfy
\begin{eqnarray} \label{recugf}
&(i)'\qquad  & \At_{K,2i} (z)-\At_{K,2i-2 }(z)= (z^2q)^{i-1}\, \Bt_{K,K-2i+2}(z)\;,\cr 
&(ii)'\qquad & \Bt_{K, 2i+1 } (z) - \Bt_{K, 2i } (z)=Ê(zq)^{2i} \, \At_{K, K-2i+\e} (zq)\;,
\cr  &(iii)'
\qquad &  \Bt_{K, 2i } (z) - \Bt_{K, 2i-1 } (z)=Ê(zq)^{2i-1} \, \At_{K, K-2i+2-\e} (zq)\;,
\,
\end{eqnarray}
 with  boundary conditions:
\beq \label{abbdry}
 \At_{K,2i}(0;q)= \At_{K,2i}(z;0)=\Bt_{K,j}(0;q) =\Bt_{K,j}(z;0) =1\;,
\eeq
and
\beq \label{bdgf}
\At_{K,0}(z)=\Bt_{K,0}(z)=0\;.
\eeq

\n {\bf Lemma 6.} The
solution to Eqs (\ref{recugf})-(\ref{bdgf}) is unique.

\n {\it Proof}: This follows from the uniqueness of the solutions of (\ref{bdryA})-(\ref{recuA}), which is itself established by  a
double induction on $n$ and $i$ (cf. sect. 7.3 in
\cite{Andr}).

The solution to Eqs (\ref{recugf})-(\ref{bdgf}) is given by the following theorem, 
whose proof is reported to the next
section.

\n {\bf Theorem 7.}
The solutions to Eqs (\ref{recugf})-(\ref{bdgf}) are
\begin{eqnarray}\label{grande}
& &\At_{K,2i}(z)= \sum_{m_1,\cdots,m_{\ka-1}=0}^\y { (-zq^{1+\e m_{\ka-1}})_\y\,
q^{N_1^2+\cdots+ N_{\ka-1}^2+L_{i}}\; z^{2N} \over (q)_{m_1}\cdots (q)_{m_{\ka-1}}
}\;,\cr
& & \Bt_{K,2i}(z)=\sum_{m_1,\cdots,m_{\ka-1}=0}^\y { (-zq^{1+\e m_{\ka-1}})_\y\,
q^{N_1^2+\cdots+ N_{\ka-1}^2+L_{i}+N}\; z^{2N} \over (q)_{m_1}\cdots (q)_{m_{\ka-1}}
}\;,\,
\end{eqnarray}
where $N_j$ and $L_j$ are defined in (\ref{defNL}) with $k$ replaced by $\kappa$ and
$\Bt_{K,2i+1}(z)$ is obtained from these expressions and $(iii)'$.
 
  Fully developed multiple $q$-series are obtained by
expanding 
$(-zq^{1+\e m_{\ka-1}})_\y$ as
\beq
(-zq^{1+\e m_{\ka-1}})_\y= \sum_{m_0=0}^\y {z^{m_0}q^{m_0(m_0+1)/2} q^{ \e m_0m_{\ka-1}}\over (q)_{m_0}}\;.\eeq

\n {\bf Corollary 8.} For $K=2\kappa$, the solutions to Eq. (\ref{recugf})-(\ref{bdgf}) reduce to 
\begin{eqnarray}\label{petite}
 & & \At_{K,2i}(z;q)= (-zq)_\y F_{\ka,i}(z^2;q) \;,\cr
& &  \Bt_{K,2i}(z;q)= (-zq)_\y F_{\ka,i}(z^2q;q)\;,
\end{eqnarray}
with $F_{\ka,i}(z^2;q)$ defined in  (\ref{ande}).

\n {\it Proof}: This follows directly from Theorem 7 with $\e=0$. An alternative direct proof, independent of Theorem 7, is given in
section 5.  See also \cite{BFJM}.



\section{Proof of Theorem 7}

The proof of (\ref{grande}) proceeds as follows (and this argument is much inspired by
\cite{Andrr}). One first rewrites the formulas (\ref{grande}) under the form
\begin{eqnarray}\label{ansatz}
& & \At_{K,2i}(z)=
\sum_{n\geq 0}{ (-zq^{1+\e n})_\y\,q^{(\ka-i)n}\, (z^2q^{n})^{(\ka-1)n}\over (q)_n}\, F_{\ka-1,i}(z^2q^{2n})\;,\cr
& & \Bt_{K,2i}(z)=
\sum_{n\geq 0}{ (-zq^{1+\e n})_\y\,q^{(2\ka-i-1)n}\, (z^2q^{n})^{(\ka-1)n}\over (q)_n}\, F_{\ka-1,i}(z^2q^{2n+1})\;\,
\end{eqnarray}
The function $\Bt_{K,2i-1}(z)$ is obtained from these expressions by
\beq \label{defb}
\Bt_{K,2i-1}(z)= \Bt_{K,2i}(z)-(zq)^{2i-1}\At_{K,K-2i+2-\e}(zq)\;.
\eeq
The function $F_{\ka,i}(z)$ is defined in (\ref{ande}) and it satisfies the recurrence relation:
\beq \label{basic}
F_{\ka,i}(z)- F_{\ka,i-1}(z) = (zq)^{i-1} F_{\ka,\ka-i+1}(zq)\;,\eeq
with boundary conditions
\beq
F_{\ka,i}(z;0)= F_{\ka,i}(0;q)= 1\qquad F_{\ka,0}(z)=F_{\ka,-1}(z)=0 \;.\eeq
Note that the vanishing of $F_{\ka,0}(z)$ together with the recurrence relation (\ref{basic}) imply that 
\beq
F_{\ka,1}(z)= F_{\ka,\ka}(zq)\;.
\eeq
The  multiple $q$-series (\ref{ande}) is the unique solution of  (\ref{basic}) with the specified boundary conditions \cite{Andrr}.

We will now show that the expressions (\ref{ansatz}) satisfy the recurrence relations (\ref{recugf}) and the boundary conditions
(\ref{abbdry}) and (\ref{bdgf}).  The latter are immediately verified: the vanishing of $F_{\ka-1,-1}(z)$ implies that of $ \At_{K,0}
(z)$ and $\Bt_{K,0}(z)$, while the precise form of (\ref{ansatz}) together with the fact that $F_{\ka,i}(z;q)$ is equal to 1 if
either $z$ or
$q$ vanishes ensure the validity of (\ref{abbdry}).

Let us first verify  the relation $(i)'$:
\beq
 \At_{K,2i} (z)-\At_{K,2i-2 }(z)=\sum_{n\geq 0}{ (-zq^{1+\e n})_\y\,q^{(\ka-i)n}\, (z^2q^{n})^{(\ka-1)n}\over (q)_n}\,
 \left[F_{\ka-1,i}(z^2q^{2n})- q^nF_{\ka-1,i-1}(z^2q^{2n})\right].\eeq
In the first step, we reorganize the square bracket as
\beq 
F_{\ka-1,i}(z^2q^{2n})-F_{\ka-1,i-1}(z^2q^{2n})+
(1-q^n)F_{\ka-1,i-1}(z^2q^{2n}) 
\eeq
and then replace the first two terms by $ (z^2q^{2n+1})^{i-1}  F_{\ka-1,\ka-i}(z^2q^{2n+1}) $ using (\ref{basic}).  That leads
to
\beq
 \At_{K,2i} (z)-\At_{K,2i-2 }(z)= R_1+R_2\eeq
with
\beq
R_1= (z^2q)^{i-1}\sum_{n\geq 0}{ (-zq^{1+\e n})_\y\,q^{(\ka+i-2)n}\,
(z^2q^{n})^{(\ka-1)n}\over (q)_n}\,
 F_{\ka-1,\ka-i}(z^2q^{2n+1}) 
\eeq
and
\beq 
R_2= \sum_{n\geq 1}{ (-zq^{1+\e n})_\y\,q^{(\ka-i)n}\, (z^2q^{n})^{(\ka-1)n}\over (q)_{n-1}}\, F_{\ka-1,i-1}(z^2q^{2n})
\eeq
(note that the summation in $R_2$ starts at $n=1$ and $(q)_n$ in the denominator has been changed to $(q)_{n-1}$ to cancel the
$(1-q^n)$ in numerator.) Let us leave $R_2$ for the moment and manipulate $R_1$. 
First write
\beq
F_{\ka-1,\ka-i}(z^2q^{2n+1})= F_{\ka-1,\ka-i+1}(z^2q^{2n+1}) - [F_{\ka-1,\ka-i+1}(z^2q^{2n+1}) - F_{\ka-1,\ka-i}(z^2q^{2n+1})]  
\eeq
and use again (\ref{basic}) to replace the last two terms by 
$-(z^2q^{2n+2})^{\ka-i} F_{\ka-1,i-1}(z^2q^{2n+2})$.
We have thus decomposed $R_1$ in two pieces:
\beq
R_1= S_1+S_2
\eeq
with 
\begin{eqnarray}
S_1&=&  (z^2q)^{i-1}\sum_{n\geq 0}{ (-zq^{1+\e n})_\y\,q^{(\ka+i-2)n}\,
(z^2q^{n})^{(\ka-1)n}\over (q)_n}\,
 F_{\ka-1,\ka-i+1}(z^2q^{2n+1}) \cr
& = &  (z^2q)^{i-1}\Bt_{K,K-2i+2+\e}(z) \
\end{eqnarray}
(to fix the second subindex of $B$ observe that $K-2i+2+\e=2(\ka-i+1)$) and
\beq
S_2=  -(z^2q)^{i-1}\sum_{n\geq 0}{ (-zq^{1+\e n})_\y\,
(z^2q^{n+2})^{(\ka-1)n+(\ka-i)}\over (q)_n}\,
 F_{\ka-1,i-1}(z^2q^{2n+2})\;.
\eeq 
Summing up our results at this point, we have
\beq \At_{K,2i} (z)-\At_{K,2i-2 }(z)= (z^2q)^{i-1}\Bt_{K,K-2i+2+\e}(z) +S_2+R_2 \;.\eeq

Let us now come back to $R_2$. We first shuffle the index $n$  to start its summation at zero:

\beq 
R_2= (z^2q)^{i-1}\sum_{n\geq 0}{ (-zq^{1+\e (n+1)})_\y (z^2q^{n+2})^{(\ka-1)n+(\ka-i)}\over
(q)_{n}}\, F_{\ka-1,i-1}(z^2q^{2n+2})\;.
\eeq
From now on, we will use the following compact  notation:
\beq
(-zq^{1+\e n})_\y f_m= \sum_{m=0}^\y {z^{m}q^{m(m+1)/2} q^{ \e m n}\over (q)_{m}} f_m\;, \eeq
i.e., we understand that $(-zq^{1+\e n})_\y$ is defined by its sum expression over $m$ so that it makes sense to insert
at its right a term that depends upon $m$.  With that notation, shifting $n$ by one unit yields:
\beq \label{idenA}
(-zq^{1+\e (n+1)})_\y= (-zq^{1+\e n})_\y \, q^{\e m}\;.
\eeq
$R_2$ reads thus
\beq 
R_2= (z^2q)^{i-1}\sum_{n\geq 0}{(-zq^{1+\e n})_\y q^{\e m} (z^2q^{n+2})^{(\ka-1)n+(\ka-i)}\over(q)_{n}}\,
 F_{\ka-1,i-1}(z^2q^{2n+2})\;.
\eeq

By comparing this expression with that of  $S_2$, we find that the summand in $R_2$ and $S_2$ are exactly the same except
for the sign and an extra factor $q^{\e m}$ in $R_2$:
\beq
S_2+R_2=
 -(z^2q)^{i-1}\sum_{n\geq 0}{(-zq^{1+\e n})_\y (z^2q^{n+2})^{(\ka-1)n+(\ka-i)}\over(q)_{n}}
(1- q^{\e m}) F_{\ka-1,i-1}(z^2q^{2n+2})\;.
\eeq
A simple observation here is that $1- q^{\e m}$ vanishes if $\e=0$. Since $\e$ can take only the values $0$ or $1$, we can thus write
\beq \label{ecase}
(1- q^{\e m})= \e (1-q^m)\;.
\eeq
$S_2+R_2$ is thus proportional to $\e$ and we can evaluate the proportionality factor at $\e=1$.
It is simple to check that  
\beq \label{idenB}
(-zq^{1+\e n})_\y (1-q^m)= zq \, (-zq^{1+\e n})_\y \, q^{m+\e n} \;.
\eeq
To be explicit: this is obtained from $(1-q^m)/(q)_m=1/(q)_{m-1}$ and by shuffling the
$m$ index in the $m$-summation.  
 Similarly, 
replacing $z\rightarrow zq$ in $ (-zq^{1+\e n})_\y$ leads to
\beq  \label{idenC}
(-zq^{2+\e n})_\y=  (-zq^{1+\e n})_\y\, q^m\;.
\eeq \label{idenD}
The comparison of the last two results gives
\beq
(-zq^{1+\e n})_\y (1-q^m)= zq\, (-zq^{2+\e n})_\y\, q^{\e n}\;.
\eeq
Substituting this into the expression of $S_2+R_2$ (and setting $\e=1$ when it appears in an exponent) leads to
\beq
S_2+R_2=-\e\, (z^{2}q)^{i-1} (zq)^{2(\ka-i)+1} \At_{K,2i-2}(zq)\;.
\eeq
Note that  we can replace $2\ka$ by $K+1$ (since $\e=1)$ in the exponent of $zq$.

Collecting all our results, we have
\begin{eqnarray}
\At_{K,2i} (z)-\At_{K,2i-2 }(z)&=& (z^2q)^{i-1} \left[\Bt_{K,K-2i+2+\e}(z) -\e \, (zq)^{K-2i+2} \At_{K,2i-2}(zq)\right]
\cr &=& (z^2q)^{i-1} \; \Bt_{K,K-2i+2}(z) \;,
\end{eqnarray}
since $\Bt_{K,K-2i+2+\e}$ is equal to $\Bt_{K,K-2i+2}$ if $\e=0$ or is given by (\ref{defb})
if
$\e=1$. We have thus completed the verification of $(i)'$.

 We now turn to the relation $(ii)'$. Note that 
the left hand side is not expressible directly in terms of a 
summand times a difference of $F$-functions due to the presence of 
$\Bt_{K,2i+1}$.  The first step amounts to reexpress it in terms of $\Bt_{K,2i+2}$:
\beq
\Bt_{K,2i+1}(z)- \Bt_{K,2i}(z)= \Bt_{K,2i+2}(z)- \Bt_{K,2i}(z)-(zq)^{2i+1} \At_{K,K-2i-\e}(zq)\;.
\eeq
Let us first concentrate on the difference between the two $\Bt$ factors:
\begin{eqnarray}
\Bt_{K,2i+2}(z)- \Bt_{K,2i}(z) &=&\sum_{n\geq 0}{ (-zq^{1+\e n})_\y\,q^{(2\ka-i-2)n}\, (z^2q^n)^{(\ka-1)n}\over (q)_n}\cr
& & \qquad \qquad \times\left[F_{\ka-1,i+1}(z^2q^{2n+1})-q^n F_{\ka-1,i}(z^2q^{2n+1})\right]\;.
\end{eqnarray}
Again, we decompose the term in square bracket as follows
\beq
[F_{\ka-1,i+1}(z^2q^{2n+1})- F_{\ka-1,i}(z^2q^{2n+1})]+ (1-q^n) F_{\ka-1,i}(z^2q^{2n+1})\;,\eeq
substitute this into the previous equation and write the corresponding two terms as $R_1'+R_2'$. 
With the identity (\ref{idenA}), $R_2'$ takes the form
\beq
R_2'= z^{2\ka-2}q^{3\ka-i-3}\sum_{n\geq 0}{ (-zq^{1+\e n})_\y\, q^{\e m} q^{(4\ka-i-4)n}\, (z^2q^n)^{(\ka-1)n}\over (q)_n}
F_{\ka-1,i}(z^2q^{2n+3})\;.
\eeq
On the other hand, $R_1'$, using (\ref{basic}), reads
\beq
R_1'= (zq)^{2i}\sum_{n\geq 0}{ (-zq^{1+\e n})_\y\,q^{(2\ka+i-2)n}\, (z^2q^n)^{(\ka-1)n}\over (q)_n} 
F_{\ka-1,\ka-i-1}(z^2q^{2n+2})\;.
\eeq
In order to demonstrate $(ii)'$, the target is to recover, within  the expression of $\Bt_{K,2i+1}(z)- \Bt_{K,2i}(z)$, that of
$(zq)^{2i}\At_{K,K-2i+\e}(zq)$, which reads (using (\ref{idenC}))
\beq
(zq)^{2i}\At_{K,K-2i+\e}(zq) = (zq)^{2i}\sum_{n\geq 0}{ (-zq^{1+\e n})_\y\,q^m q^{(2\ka+i-2)n}\, (z^2q^n)^{(\ka-1)n}\over (q)_n} 
F_{\ka-1,\ka-i}(z^2q^{2n+2})\;.
\eeq
Apart from the factor of $q^m$ and the value of the second index of the function $F$, the last two expressions are identical. This
indicates the way we should manipulate $R_1'$. First write
\beq
F_{\ka-1,\ka-i-1}(z^2q^{2n+2})= F_{\ka-1,\ka-i}(z^2q^{2n+2}) - 
[F_{\ka-1,\ka-i}(z^2q^{2n+2})- F_{\ka-1,\ka-i-1}(z^2q^{2n+2})]\;.
\eeq
This decomposes $R'_1$ in two pieces
$S_1'+S_2'$ with 
\beq
S_1'= (zq)^{2i}\sum_{n\geq 0}{ (-zq^{1+\e n})_\y\,q^{(2\ka+i-2)n}\, (z^2q^n)^{(\ka-1)n}\over (q)_n} 
F_{\ka-1,\ka-i}(z^2q^{2n+2})
\eeq
and (using again (\ref{basic}))
\beq
S_2'=-z^{2\ka-2}q^{3\ka-i-3}\sum_{n\geq 0}{ (-zq^{1+\e n})_\y\,  q^{(4\ka-i-4)n}\, (z^2q^n)^{(\ka-1)n}\over (q)_n}
F_{\ka-1,i}(z^2q^{2n+3})\;.
\eeq 
In $S_1'$, we then insert a factor $q^m$ as follows: $1= q^m+(1-q^m)$ and write the resulting two contributions as
\beq
S_1'= (zq)^{2i}\At_{K,K-2i+\e}(zq) + T_2'\eeq
and (with (\ref{idenB})):
\beq 
T_2'= (zq)^{2i+1}\sum_{n\geq 0}{ (-zq^{1+\e n})_\y\, q^{\e n+ m} q^{(2\ka+i-2)n}\,(z^2q^n)^{(\ka-1)n}\over (q)_n}
F_{\ka-1,\ka-i}(z^2q^{2n+2})\;.\eeq
Collecting the results of this paragraph, we see that to  complete the proof of $(ii)'$ we only have to show that
\beq \label{reste} R_2'+S_2'+T_2'-(zq)^{2i+1}\At_{K,K-2i-\e}(zq)=0\;.\eeq
By comparing $R'_2$ and $S'_2$, we notice that their summands are identical, up to the sign and to an extra $q^{\e m}$ in $R_2'$.
$R'_2+S_2'$ contains thus the factor $(1-q^{\e m})$ which can be handled as previously (cf. eqs (\ref{ecase}) and
(\ref{idenB})). The result is
\beq \label{rets}
R'_2+S_2'= -\e\,  z^{2\ka-1}q^{3\ka-i-2}\sum_{n\geq 0}{ (-zq^{1+\e n})_\y\, q^{ m} q^{(4\ka-i-3)n}\, (z^2q^n)^{(\ka-1)n}\over
(q)_n} F_{\ka-1,i}(z^2q^{2n+3})\;.\eeq
Combining next $T_2'$ with $-(zq)^{2i+1}\At_{K,K-2i-\e}(zq)$ leads to
\begin{eqnarray} \label{tetA}
T_2'-(zq)^{2i+1}\At_{K,K-2i-\e}(zq)&=& (zq)^{2i+1}\sum_{n\geq 0}{ (-zq^{1+\e n})_\y\, q^{ m} q^{(2\ka+i-2+\e)n}\,
(z^2q^n)^{(\ka-1)n}\over (q)_n} \cr
& & \quad \times [F_{\ka-1,\ka-i}(z^2q^{2n+2})- F_{\ka-1,\ka-i-\e}(z^2q^{2n+2})]\;.\
\end{eqnarray}
Using (\ref{basic}) in a slightly modified form, i.e., as
\beq
F_{\ka-1,\ka-i}(z^2q^{2n+2})- F_{\ka-1,\ka-i-\e}(z^2q^{2n+2})= \e\,  (z^2 q^{2n+3})^{(\ka-i-1)}
F_{\ka-1,i}(z^2q^{2n+3})\;,
\eeq
we are led to
\beq
T_2'-(zq)^{2i+1}\At_{K,K-2i-\e}(zq)= -(R_2'+S_2')\;,\eeq
which demonstrates (\ref{reste}) and thus $(ii)'$.

Finally, with $\Bt_{K,2i-1}(z)$ defined by (\ref{defb}), relation $(iii)'$ is an identity.  We have thus completed the
proof of the relations (\ref{ansatz}) or equivalently of Theorem 7.

\section{Two simple applications to partition counting}

By adding  the staircase $(m-1,m-2,\cdots,1,0)$ to the vector $(n_1,\cdots n_m)$, we transform it into an
ordinary partition. With $\la_j=n_j+m-j$, the weakly decreasing conditions (\ref{oror}) become
\beq \label{parc}
\la_j\geq \la_{j+1} \qquad {\rm and} \qquad \la_j\geq \la_{j+2}+2\;,
\eeq
while the $K$-restrictions (\ref{allo}) take the form
\beq \label{pard}
\la_j \geq  \la_{j+K-1} +K  \qquad{\rm or} \qquad  \la_j = \la_{j+1}= \la_{j+K-2} +K-1=  \la_{j+K-1} +K-1\;.\eeq
To transform a generating function for $K$-restricted jagged partitions to one for partitions subject to (\ref{parc}) and
(\ref{pard}), we simply need to replace $z^N$ by $z^N q^{N(N-1)/2}$. Two limiting cases of our general result are of interest.

\n {\bf Corollary 9.} The number of partitions satisfying $\la_j\geq \la_{j+2}+2$ is given by
\beq \label{libre}
\lim_{\kappa\rw\y}\sum_{m,n\geq 0} A_{2\ka,2\ka}(m,n) z^mq^{m(m-1)/2+n}= \sum_{m_0,m_1\geq 0}
{q^{(m_0+m_1)^2+m_1^2}z^{m_0+2m_1}\over (q)_{m_0}(q)_{m_1} }= F_{3,3}(z)\;.
\eeq 

\n {\it Proof}: In the limit $\kappa\rw\y$, the restrictions can be disregarded and we are left with unrestricted jagged partitions,
which, by Corollary 8, satisfy
\beq \label{gefc}
\lim_{\kappa\rw\y}\A_{2\ka,2\ka}(z)= (-zq)_\y \lim_{\kappa\rw\y} F_{\ka,\ka}(z^2) = {(-zq)_\y\over
(z^2q)_\y}\;.
\eeq
The second equality follows from Theorem 2 of \cite{Andrr}. 
By expanding $(-zq)_\y/ (z^2q)_\y$ and replacing $z^N$ by $z^N q^{N(N-1)/2}$, we recover the first equality of (\ref{libre}), which is
seen to be equivalent to  $F_{3,3}(z)$, as it should (since we simply recover a special case of \cite{Andrr} quoted in the
introduction).

\n {\bf Corollary 10.} The number of partitions satisfying $\la_j\geq \la_{j+2}+3$ is given by
\beq \label{castrois}
\sum_{m,n\geq 0} A_{3,4}(m,n) z^mq^{m(m-1)/2+n}=
\sum_{m_0,m_1\geq0} {q^{m_0^2+3m_1(m_0+m_1-1/3)} z^{m_0+2m_1}\over (q)_{m_0}(q)_{m_1} } \;.\eeq

\n {\it Proof:} For $K=3$, all the restrictions on the  partitions defined by (\ref{parc}) and (\ref{pard}) reduce to 
\beq
\la_j\geq \la_{j+2}+3 \;.\eeq
We can apply (\ref{grande})  to the counting of such partitions by considering
$\A_{3,4}(z,q)$ (i.e.,
setting $i=2$ to take into account all boundary conditions). Replacing again $z^N$ by $z^Nq^{N(N-1)/2}$, leads to 
(\ref{castrois}).

This provides a quite simple derivation of a specialization of the generating function of partitions with difference 3
at distance $k-1$: $\la_i\geq \la_{i+k-1}+3$ obtained in \cite{Mu1} (cf. their Eq
6). (See also Theorem 9.9 of \cite{Mu2} for the generating functions of the  restricted partitions
for this generic case (all $k$). Our result is also a specialization of the one presented in Theorem 5.14 of \cite{Mu3}
pertaining to the case $\la_i\geq \la_{i+2}+\ell.$)

\section{Product form of the specialized generating function}

Let us return to the general multiple sum $\At_{K,2i}(z)=\At_{K,2i}(z;q) $. For $z=1$, it can be regarded
as the sum-side of a generalized version of the Rogers-Ramanujan identities. In this section, we
display the corresponding  product form together with its combinatorial interpretation.

\n {\bf Theorem 11.} The product form of ${\tilde A}_{K,2i}(1;q)$, with $K=2\kappa-\e$ and $1\leq i\leq
\kappa$  reads
\begin{eqnarray} \label{theor}{\tilde A}_{K,2i}(1;q)&=& \prod_{n=1}^\infty  (1+ q^n)  \prod_{n\not
= 0,
\pm i
\;{\rm mod}\, (K +1)}^\infty  (1- q^n)^{-1} \qquad (\epsilon=0,1,; i<(K+1)/2 )\cr
 &=& 
 \prod_{n\not = 0\;{\rm mod}\;
\kappa}^\infty [(1+ q^n) (1- q^n)^{-1}] \qquad (\epsilon=1, i= \kappa)\;.
\end{eqnarray}

\n Proof: Using the simple identity $(-q^{1+\e m_{\kappa-1}})_\y= (-q)_\y / (-q)_{\e m_{\kappa-1}}$, we
can rewrite $\At_{K,2i}$, given by  (\ref{grande}), as
\beq \label{idfr}
\At_{K,2i}(1;q)= (-q)_\y\sum_{m_1,\cdots,m_{\ka-1}=0}^\y {
q^{N_1^2+\cdots+ N_{\ka-1}^2+L_{i}} \over (q)_{m_1}\cdots (q)_{m_{\ka-2}}
(q^{2-\e};q^{2-\e})_{m_{\ka-1}}}\;.
\eeq
Up to the prefactor $(-q)_\y$, the multiple sum is now in a form equivalent to one used in \cite{Bres}. The result
(\ref{theor}) for $i<(K+1)/2 $ follows directly from Theorem 1 of \cite{Bres}. It only remains to consider the case
where
$\e=1$ and $i= (K+1)/2=\kappa$.  But this  is implicitly treated in Lemma 1 of \cite{Bres}, which leads immediately
to the second line of (\ref{theor}). (In that case, the restriction $n\not = 0,
\pm\kappa\;{\rm mod}\; 2\kappa$ reduces to $n\not = 0\;{\rm
mod}\; \kappa$.)

Manifestly, in all cases but $\e=1,\, i=\kappa$, the factor $(-q)_\y$ can be dropped from both sides of (\ref{theor})
(cf.  (\ref{idfr}) for the left hand side). By doing so, we recover the Andrews-Gordon identities ($\e=0)$ \cite{An} and
the Bressoud identities $(\e=1)$
\cite{Bres}. For
$\e=1,\, i=\kappa$, (\ref{theor}) appears to be a new identity.

  Note that for $\e=1,
i<\kappa$, we have the following expression:
\beq{\tilde A}_{2\kappa-1,2i}(1;q)= {(-q)_{\infty}\over (q)_{\infty}}( q^i,
q^{2\kappa-i},q^{2\kappa}; q^{2\kappa})_{\infty}\;.\eeq
For $i=1$, this is   equal to $ F_{\kappa, 1} (-1/q;1;q)$ (cf.  Lem. 2.6 of \cite{Lov}), a specialization of the
function $F_{\kappa, i} (a;z;q)$ of Andrews
\cite{Andrr,Andr} ($F_{\kappa, i} (z;q)$ in (\ref{ande}) being its $a=0$
version).   
  For
$i=\kappa$, we have
\beq{\tilde A}_{2\kappa-1,2\kappa}(1,q)= {(-q)_{\infty}\over (q)_{\infty}}
{(q^\kappa;q^\kappa)_{\infty}\over(-q^\kappa;q^\kappa)_{\infty}}
= F_{\kappa,\kappa}(-1;1;q)\;,\eeq
 where the last identity is proved
in \cite{Lov}, Lem. 2.5. For $i=2, K=3$, this is the product side of Lebesgue's identity 
 (cf. \cite{Andr} Cor. 2.7 with
 $a=1$).

The combinatorial interpretation of the Rogers-Ramanujan-type identities (\ref{theor}) (with $\At_{K,2i}$ given by
(\ref{grande})) relies on the description of jagged partitions as overpartitions.
 Recall that an overpartition is a partition in which the
first occurence of a number may be overlined \cite{CoLo}.
An overpartition is thus equivalent to a pair $(\bar{\alpha},\beta)$ of
partitions with the constraint that the parts of $\bar{\alpha}$ are distinct (i.e., they are the overlined parts).
There is a natural bijection between overpartitions and jagged
partitions, obtained as follows \cite{Lo}. Replace adjacent integers $(n,n+1)$ within
the jagged partition by $2n+1$ and similarly replace adjacent integers
$(n,n)$ by $2n$. The numbers thus obtained form the parts of $\beta$. The remaining entries
of the jagged partitions are necessarily non-zero and distinct integers; they
build up $\bar{\alpha}$. On the other hand, given an overpartition
$(\bar{\alpha}, \beta)$, one first decomposes all entries of $\beta$ according
to their parity, either as $2n=(n,n)$ or $2n+1=(n,n+1)$ and uses the
resulting (adjacent) parts together with those of $\bar{\alpha}$, to construct a jagged
partition according to the restrictions (\ref{oror}). This is unique and this demonstrates the bijective character
of the correspondence. (Observe that the equivalence between the set of jagged partitions of weight $n$ and  pairs
of partitions
$(\bar{\alpha},\beta)$ whose weights add up to $n$, is a direct consequence of the generating function
(\ref{gefc})). 

The above bijection and Theorem 11 lead directly to the following.

\n {\bf Corollary 12.} The number $A_{K,2i}(n)$ (with $1\leq i\leq [(K+1)/2]$ and $K=2\kappa-\e$) of jagged
partitions of weight
$n$ satisfying the
 restrictions 
(\ref{allo}) and containing at most $i-1$ pairs 01 is equal to the number of overpartitions $(\bar{\alpha},\beta)$
of combined weight $n$  where parts of $\beta$ are not equal to
$0,\pm i$ mod
$ K+1$ if
$1\leq i< (K+1)/2$ or  where both parts  of $\bar{\alpha}$ and $\beta$ are not equal to 0 mod $ \kappa$ if $\e=1$ and
$i=\kappa.$

A completely different combinatorial interpretation of the identity (\ref{theor}) (without the $(-q)_\y$ factor) is
given in 
\cite{An, Bres} (except for the case
$\e=1$ and
$i=\kappa$ which is not covered  in \cite{Bres}).  We stress that by including a $(-q)_\y$ factor, we end up with a new
combinatorial description of these previously known identities.  Note also that the sum-side does not seem to have a
natural interpretation in terms of overpartitions with a difference conditions at distance $K-1$.

\section{Complementary remarks}

Before concluding, we would like to present some clarifying remarks.  The first one concerns the relationship between the
$K$-restrictions (\ref{allo}) and the recurrence relations (\ref{recuA}).  A basic observation is that they have a {\it dual} role:
the restriction conditions specify the {\it allowed} jagged partitions, while the recurrence relations are controlled by
the {\it excluded} jagged partitions.  Hence, if at first sight it might not seem natural to have a restriction
formulated in terms of an `or'-type condition, it is clear that the introduction of an alternative allows for more jagged
partitions than with a single restriction.  And this implies that there are  less excluded jagged partitions, meaning, in turn, 
 that the recurrence relations are simplified. In other words,  if the restriction was formulated in terms of a single step-$K$
difference-one condition,
that  would result in a system of recurrence
relations more complicated  than (\ref{recuA}) and  unlikely to be solvable in closed form.

To make the duality more explicit, observe that the $K$-restrictions (\ref{allo})
are equivalent to excluding all jagged partitions containing a $K$-component subvector $(n_j,\cdots , n_{j+K-1})$ of either one of the
following form:
\beq \label{exclus}
(\underbrace{p,\cdots,p}_{K-2\ell},\underbrace{p-1,p,\cdots,p-1,p}_{2\ell}) \qquad {\rm or }\qquad
(\underbrace{p,p+1,\cdots,p,p+1}_{2\ell},\underbrace{p,\cdots,p}_{K-2\ell}) 
\eeq
with $0\leq \ell\leq [K/2]$. When viewed from this angle, the naturalness of the condition (\ref{allo}) reveals itself: it amounts to
exclude precisely one  subvector of length $K$ for each value of the weight $n= \sum_{r=j}^{K-j+1} n_r$ ($n\geq  [(K+1)/2]$).  But
this pattern of excluded subvectors pops up directly from the recurrence relations. The condition $(i)$ indicates that we need to
exclude all vectors whose tail is of the form
\beq \label{exclua}
(\cdots,\, \underbrace{1,\cdots,1}_{K-2\ell},\underbrace{0,1,\cdots,0,1}_{2\ell}) \;,\eeq
(with $\ell=i-1$), while the conditions $(ii)$ and $(iii)$ amounts to eliminate all vectors with the following tail:
\beq \label{exclub}
(\cdots,\, \underbrace{1,2,\cdots,1,2}_{2\ell},\underbrace{1,\cdots,1}_{K-2\ell}) 
\eeq
(with $\ell=\ka-i$ or $\ka-i+1-\e$ respectively). Since the restriction conditions are invariant under a shift of all the parts
$n_j$  by the same integer
$p$ (i.e., by adding
$(p^m)$ to the jagged partition), the exclusion condition must share this invariance property; this lifts the tail-exclusions
(\ref{exclua}) and (\ref{exclub}) to the general ones (\ref{exclus}).

Next, since our proof of Theorem 7 is not constructive and based on the judicious ansatz (\ref{grande}), it is fair ot
present some rationale underlying this ansatz. At first, we already knew from
\cite{BFJM} that for
$K=2\ka$, 
\beq 
\At_{K,2i}(z)= (-zq)_\y F_{\ka,i}(z^2)\;.
\eeq
Here is a very quick proof, independent of Theorem 7.  Set  \beq
\At_{K,2i}(z)= f(z) F_{\ka,i}(z^2)
\eeq
and substitute this into $(i)'$, using ({\ref{basic}); this leads to
\beq
\Bt_{K,2i}(z)= f(z) F_{\ka,i}(z^2q)\;.
\eeq
Then from $(iii)'$ we get
\beq
\Bt_{K,2i-1}(z)= f(z;q) F_{\ka,i}(z^2q)- z^{2i-1}q^{2i-1}f(zq) F_{\ka,\ka-i+1}(z^2q^2)\;.
\eeq
The substitution of these expressions into $(ii)'$ yields then
\beq f(z;q)= (1+zq)f(zq) \qquad \Rightarrow\qquad f(z)= (-zq)_\y\;.
\eeq
In that case, the $m_0$ mode defined by the sum expression  of $(-zq)_\y$ is thus independent of the $m_j$ ones
of
$F_{\ka,i}$. Given that in the large-$K$ limit, the parity of $K$ should not  matter anymore, we should recover the above simple
result even for
$K$ odd, as $K\rw\y$. This means that if the
$m_0$ mode is coupled to some other modes $m_j$, when $K$ is odd, this coupling should disappear as $K\rw\y$. In the
multiple-sum expression (\ref{usual}) of
$F_{\ka,i}$, we see that there are  terms like $q^{jm_{j}^2}$, so that in the large
$K$ (or $\kappa$) limit, the only contributing values of the modes $m_j$ with $j$ of the order of $\ka$ are $m_j=0$ (with the usual
assumption that $q<1$). From these considerations, we thus knew that $m_0$ could couple only with those modes $m_j$ with $j$ of
the order of $\ka$. The natural guess is to look for a single coupling with the mode with largest subindex, $m_{\ka-1}$. This is also
a very natural hypothesis if we expect  an iterative formula like (\ref{ansatz}) to  exist (where the iteration is on $\ka$) in which
the dependence upon the modes $m_j$,
$1\leq j\leq \ka-2$ is factored out.

\section{Conclusion}

We have presented a rather interesting extension of the generating function  counting  partitions with 
difference 2 at distance $k-1$, by enumerating novel types of partitions (dubbed `jagged') subject to a new type of
restriction. That the rather complicated  restriction considered here (dictated, as already pointed
out, by a physical problem) leads to a set of $q$-difference equations  solvable by functions so similar to
the original Andrews' multiple sums is certainly quite remarkable.  On the other hand, 
 there is
 a whole hierarchy of jagged partitions generalizing those considered here, whose further study appear to be quite interesting.
Preliminary results along that direction are presented in \cite{FJM}.

\vskip0.3cm
\noindent {\bf ACKNOWLEDGMENTS}

We thank J. Lovejoy for communicating to us the bijection between overpartitions and jagged partitions
and also for pointing out reference \cite{Bres}.

\end{document}